\documentclass[aps,prl,twocolumn,showpacs,superscriptaddress,floatfix]{revtex4}
\usepackage{graphicx,graphics}
\usepackage{dcolumn}
\usepackage{amsmath,amssymb,amsfonts}
\usepackage{latexsym,verbatim}
\usepackage{bm}
\usepackage{color}
\usepackage{ulem}
\usepackage[breaklinks=true,colorlinks,citecolor=blue,linkcolor=blue,urlcolor=blue]{hyperref}
\def\up{\uparrow}
\def\down{\downarrow}
\def\bef{\begin{framed}}
\def\eef{\end{framed}}
\def\be{\begin{equation}}
\def\ee{\end{equation}}
\def\ber{\begin{eqnarray}}
\def\eer{\end{eqnarray}}

\def\sigmabold{\mbox{\boldmath $\sigma$}}
\def\rv{{\bf r}}
\def\zv{{\bf \hat z}}
\def\pv{{\bf p}}

\def\vv{{\bf v}}

\def\jv{{\bf j}}

\def\Sv{{\bf S}}

\def\vv{{\bf v}}
\def\nn{\nonumber}

\begin{document}
\title{Current-induced spin polarization at the surface of metallic films: a theorem and an {\it ab initio} calculation}
\author{I. V. Tokatly}
\affiliation{%
Departamento de F\'{\i}sica de Materiales, 
Universidad del Pais Vasco UPV/EHU, 20080 San Sebasti\'an/Donostia, Spain}
\affiliation{IKERBASQUE, Basque Foundation for Science, E-48013 Bilbao, Spain}

\author{E. E. Krasovskii}
\affiliation{%
Departamento de F\'{\i}sica de Materiales, 
Universidad del Pais Vasco UPV/EHU, 20080 San Sebasti\'an/Donostia, Spain}
\affiliation{Donostia International Physics Center (DIPC), E-20018 San Sebasti\'an, Spain}	
\affiliation{IKERBASQUE, Basque Foundation for Science, E-48013 Bilbao, Spain}

\author{Giovanni Vignale}
\email{vignaleg@missouri.edu}
\affiliation{Department of Physics and Astronomy, University of Missouri, Columbia, Missouri 65211, USA}
\affiliation{Donostia International Physics Center (DIPC), E-20018 San Sebasti\'an, Spain}	

\begin{abstract}
 The broken inversion symmetry at the surface of a metallic film (or,
 more generally, at the interface between a metallic film and a
 different metallic or insulating material) greatly amplifies the
 influence of the spin-orbit interaction on the surface properties.
 The best known manifestation of this effect is the momentum-dependent
 splitting of the surface state energies (Rashba effect).  Here we
 show that the same interaction also generates a spin-polarization of
 the bulk states when an electric current is driven through the bulk
 of the film. For a semi-infinite jellium model, 
which is representative of metals with a closed Fermi surface,
 we prove as a theorem that, regardless of the shape of the
 confinement potential, the induced surface spin density at each
 surface is given by $\Sv =-\gamma \hbar \zv \times \jv$, where $\jv$ is the
 particle current density in the bulk, $\zv$ the unit vector normal to
 the surface, and $\gamma=\frac{\hbar}{4mc^2}$ contains only fundamental constants.
For a general metallic solid $\gamma$ becomes a material-specific parameter 
that controls the strength of the interfacial spin-orbit coupling.  
Our theorem, combined  with an {\it ab initio} calculation of the spin
 polarization of the current-carrying film, enables a
determination of $\gamma$, which should be useful in modeling the spin-dependent scattering
of quasiparticles at the interface.
 \end{abstract}
%
\maketitle
\section{Introduction}

Physical phenomena in which  an electric current is converted into a spin polarization and/or a spin current, are receiving a great deal of attention in the context of orbital spintronics\cite{Dyakonov71,Lyanda-Geller89,Edelstein1990,Hirsch99, Zhang00, Murakami03, Sinova04, Handbook,
Raimondi_2Dgas_PRB06, Culcer_SteadyState_PRB07, Culcer_Generation_PRL07, Culcer_SideJump_PRB10, TsePRB05,
GalitskiPRB06, Tanaka_NJP09,Hankiewicz09,TenYears2010} -- an appealing alternative to  ``classical" spintronics\cite{Baibich1988,Grunberg1986,Fert93,Wolf01,rmp_76_323}.  While in classical spintronics the spin dynamics is mainly controlled by exchange interactions, in orbital spintronics the central role is played by the spin-orbit (SO) interaction, which allows direct manipulation of the spins by electric fields.\cite{Valenzuela_Nat06,Takahashi_Revese_PRL07,Takahashi_GSH_NatMater08,Ioan2010,Kato04,Kato04-2,Sih05,Inoue2003,Yang2006,Chang2007,Koehl2009,Kuhlen2012}   In recent years both the exchange interaction-based approach and the spin-orbit interaction-based one have been shown to be viable for achieving  current-induced switching of the magnetization of a ferromagnetic metal.\cite{Ioan2010,Ralph_FMR_SHE_11,Ralph_MgnSwc_SHE_11,Ralph_Pt_SHE_Rvw_11}   Although spin-orbit interactions are generally much weaker than exchange interactions, 
they are known to produce a characteristic linear in
momentum spin splitting of surface states -- the 
so-called Rashba effect\cite{Rashba84}, which is observed in semiconductor as
well as metallic interfaces.  The size of this
splitting can be tuned by an external electric field, which creates
the  possibility of using the effect as the basis for a field-effect transistor.\cite{DattaDas}  In a different
manifestation of the Rashba effect, a non-equilibrium spin
accumulation in the surface states can produce a spin-galvanic current
(or voltage) -- a phenomenon that has been experimentally demonstrated
in semiconductors\cite{Ganichev2002} and, more recently, in metallic
(Bi/Ag) interfaces.\cite{Rojas-Sanchez2013}
 
The Rashba splitting of surface states is by no means the only
important manifestation of SO at a surface.  Recently, it has been
pointed out that the surface-induced SO coupling can have large
effects also on the bulk states~\cite{Kimura2010,Krasovskii2011,Wang2013}
of a macroscopic thin metallic film
that is sandwiched between two insulating barriers.  It has been
predicted that such films could exhibit large spin Hall angles, and
that an electric polarization perpendicular to the surface should
appear to second order in the electric field driving a current in the
plane of the film.\cite{Wang2013}

\begin{figure}\label{MetalFilm}
\begin{center}
\includegraphics[width=0.80\columnwidth]{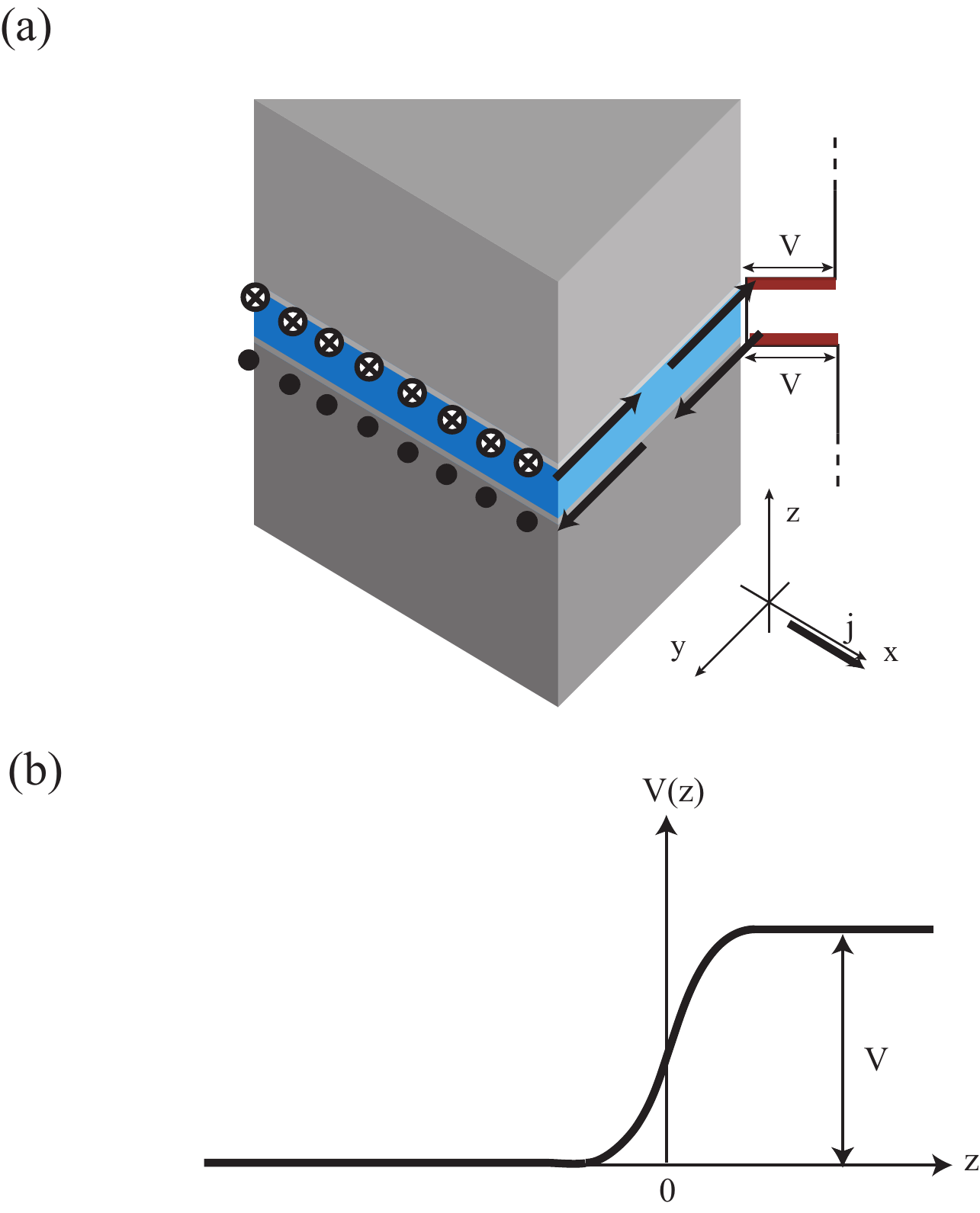}
\caption{(a) A metal film sandwiched between two insulators and separated from them by potential barriers of height $V$.  The black arrows and dots indicate the direction of the spin polarization induced by the interfacial spin-orbit interaction when a current $\jv$ flows along the $x$ axis  in the bulk of the film.
(b) A more detailed view of the confining potential as a function of the coordinate $z$ perpendicular to the interface.  Only one interface is shown in this drawing,  at $z=0$, the other one being located far away on the negative $z$ axis.}
\end{center}
\end{figure}

In this paper we focus on the non-equilibrium spin polarization that
appears in the immediate vicinity of the surface of a metal when a
uniform current is driven throughout the bulk of the metal, parallel to the surface.
The effect bears some similarity to the well known Edelstein
effect,\cite{Lyanda-Geller89,Edelstein1990} which occurs in two-dimensional electron
gases at the surface of semiconductors and metals.   The crucial
difference here is that the spin polarization of interest occurs in a continuum of bulk states scattering off the surface, with a relatively minor contribution from the surface state.  
The induced spin polarization is
perpendicular to the current, parallel to the surface, and confined to
a distance of order $k_F^{-1}$ of the surface, where $k_F$ is the
three-dimensional Fermi wave vector of the bulk electrons.  For a
semi-infinite jellium model, taken to be representative of metals with
a simply connected Fermi surface, we prove as a theorem that the
surface spin density (i.e., the spin density integrated over the
coordinate perpendicular to the surface) is completely independent of
the details of the confinement potential and is given by the elegant
formula
\be\label{MainResult}
\Sv =-\gamma \hbar \zv \times \jv\,,
\ee
where $\jv$ is the particle current density in the bulk, $\zv$ the
unit vector normal to the surface, and $\gamma=\frac{\hbar}{4mc^2}$ (with $m$ the electron mass) 
contains only fundamental constants.  The relation holds also for a general metallic solid, 
but $\gamma$ then becomes a material-specific parameter
that controls the strength of the interfacial spin-orbit coupling.
In a thin macroscopic metal film having two surfaces separated by a distance much larger than $k_F^{-1}$ the two surfaces induce independent spin polarizations of  opposite sign, such that the total integral of the spin density across the film vanishes.  This situation can be described as a kind of bulk spin Hall effect, in which the current flowing in the bulk of the film induces spin accumulations of opposite signs on the two surfaces.  
But the driving force is not the spin-orbit interaction in the bulk of
the metal, nor the spin-orbit interaction with impurities -- rather,
the spin-orbit interaction with the surface confinement potential, if 
one adopts the  jellium model description, or, alternatively,  if one considers the real material, the combination of the atomic spin-orbit interaction with the loss of inversion 
symmetry induced by the termination of the bulk crystal.
{\bf This kind of spin Hall effect is  different
from the  one investigated in Ref.~\onlinecite{Wang2013} (where 
the accumulated spin was perpendicular to the
plane of the film),   but is  quite similar (although conceptually distinct from it) to the standard spin-Hall effect, which arises from the spin-orbit interaction in the bulk of the metal.}  It is remarkable
that the final result has the simple and universal form of Eq. (1):
such a structure is  reminiscent of exact results about
impurities in metals, where the summation of contribution from all
occupied scattering states generates the Friedel sum rule or Fumi's
theorem.\cite{Mahan2000b}.  A similar result for normal-metal-superconductor interfaces was obtained by Edelstein.\cite{Edelstein2003}
    
There remains the fundamental problem of determining the value of the constant $\gamma$, which mimics 
in the jellium model the surface spin-orbit coupling of the real material.  
This constant is expressed in 
terms of the effective electron mass $m$ and the effective Compton wavelength $\lambda_c$ as follows:
 \be\label{gammalambda}
 \gamma = s \frac{m \lambda_c^2}{4 \hbar}\,,
 \ee
where $s =\pm1$ is the overall sign of the expression (for a free
electron in vacuum one has $s=+1$ and $\lambda_c = \hbar/mc\simeq
10^{-2}$\AA, but these values can be dramatically different in a solid
state environment: for electrons in GaAs, for instance, $s=-1$ and
$\lambda_c \simeq 2$\AA) To determine the quantities $s$ and
$\lambda_c$ in any specific situation one must 
draw on detailed microscopic calculations, which take into
account the effect of the atomic spin-orbit
interaction on the electronic states.  Here we propose a novel
approach to the calculation of $s$ and $\lambda_c$, based on the use
of Eq. (1).  The idea is to perform an {\it ab initio} calculation of
the spin polarization of the bulk states of a thin metal film in the
presence of a homogeneous current.  Assuming the standard relaxation
time approximation the current is introduced by shifting the Fermi
surface in momentum space.
The resulting coefficient of proportionality between the surface spin density and the particle current density yields an {\it ab initio} estimate of $\gamma$.  
In what follows, we apply this idea to the calculation of $\gamma$ at the surface of a gold film.  
Although in Au(111) there is a Rashba-split surface state 
at the Fermi level, in our {\it ab initio} calculation we find its
contribution to the surface spin polarization to be an order of 
magnitude smaller than the contribution from the bulk continuum.  
Moreover, a careful study of the three-dimensional spin density confirms that the
induced spin is confined to a relatively small region ($\sim
k_F^{-1}$) near the surfaces.  We believe that Eq. (1), combined with
{\it ab initio} theory, provides a remarkably simple approach to the
determination of $\gamma$, a crucial parameter for the spintronics of
thin metal films.  

{\it Proof of the theorem} -- 
We consider the setup of Fig. 1: a semi-infinite three-dimensional electron gas (jellium model) is confined to the half space with $z<0$ by a potential that rises from $V(z)=0$ for $z \to -\infty$ to $V(z)=V$ for $z \to \infty$.  No assumption is made about the  shape of this potential.
It is further assumed that the chemical potential of the electrons $\mu$ is smaller than the barrier height $V$.   Our objective is to calculate the integrated spin density induced by the interface in the infinite jellium.  In the absence of spin-orbit interaction the electronic states are characterized by a conserved two-dimensional momentum $\pv$ in the x-y plane (parallel to the interface) and by an asymptotic one-dimensional wave vector $k>0$ in the z-direction (perpendicular to the interface):
\be
\psi_{\pv,k}(\rv,z)=  e^{i\pv\cdot\rv} \varphi_{k}(z)\,,
\ee  
where the wave functions $\varphi_{k}(z)$ are spinors of definite spin orientation ($\up$ or $\down$) and  have the asymptotic form 
\be
\varphi_{k}(z) = \left\{\begin{array}{c}
 e^{ikz}+\hat r_{k}e^{-ikz}\,,~~z\to-\infty\\ 
 (1+\hat r_{k})e^{-\kappa z}\,,~~~z\to +\infty    
\end{array}\right.
\ee
where $\hat r_k$ is a phase factor and $\kappa = \sqrt{2mV-k^2}$.  

This classification of states is essentially preserved by the spin-orbit interaction of form
\be\label{HSO}
H_{SO}(z) = \gamma V'(z)(\zv \times \vv_\pv)\cdot\sigmabold\,,
\ee
where $\vv_\pv$ is the velocity operator and $V'(z)$ is the derivative of the potential with respect to $z$.   The only difference is that $\hat r_k$ becomes a
unimodular $2\times 2$ matrix, mixing $\up$ and $\down$ spin states.

The spin polarization  at position $z$ is obtained from the trace of the spectral function
\be
A_\pv(z,\omega) = -2 \Im m G_\pv^R(z,z,\omega)\,,
\ee 
where the retarded Green function  $G_\pv^R(z,z',\omega)$ is a  $2 \times 2$ matrix in spin space, in the following manner 
 \be
{\bf s}(z) = \frac{1}{2} \sum_{\pv} {\rm Tr}\int_{-\infty}^{+\infty} \frac{d\omega}{2\pi} f_\pv(\omega) \left[\sigmabold A_\pv(z,\omega) \right]\,,
\ee
where $f_\pv(\omega)$ is the average occupation of a state of  parallel momentum $\pv$ at energy $\omega$.  In equilibrium this would be the Fermi distribution at chemical potential $\mu$ and temperature $T$, 
$f(\omega)=[e^{\beta(\omega-\mu)}+1]^{-1}$ independent of $\pv$.   In a current-carrying state, such as we are considering here, the occupation is given by a displaced Fermi distribution function
$f_\pv(\omega) = f(\omega-\pv\cdot \vv_d)$, 
where $\vv_d$ is the average drift velocity of the electrons in the plane of the film.
The surface spin density ${\bf S}$, defined as  ${\bf s}(z)$ integrated over $z$ is then given by  
\be\label{SSD}
\Sv = - \sum_{\pv} \int_{-\infty}^{+\infty} \frac{d\omega}{2\pi} f_\pv(\omega)  \Im m \int dz {\rm Tr} \left[\sigmabold G^R_\pv(z,z,\omega) \right]\,.
\ee
Notice that we have set $\hbar=1$ in these calculations.

This formula is exact and obviously yields zero spin polarization if spin-orbit coupling is absent.  We will now proceed to evaluate $G^R_\pv(z,z,\omega)$ to first order in the strength of spin-orbit coupling.   The first-order expression for this is
\be
G^R_\pv(z,z,\omega) = \int dz'  g^R_\pv(z,z',\omega)  H_{SO}(z') g^R_\pv(z',z,\omega)
\ee
where $H_{SO}(z)$ is defined in Eq.~(\ref{HSO})  and $g^R_\pv(z,z',\omega)$ is the retarded Green's function in the absence of spin-orbit coupling, i.e.,
\be
 g^R_\pv(z,z',\omega)= \sum_k \frac{\varphi_k(z)\varphi_k^*(z')}{\omega - \epsilon_\pv(k)+i\eta}\,,
 \ee
 where $ \epsilon_\pv(k) \equiv \frac{p^2}{2m_\parallel}+\frac{k^2}{2m}$.  Making use of the explicit form of $H_{SO}(z)$ we  find
\ber\label{Base}
&&\Im m\int dz {\rm Tr} \left[\sigmabold G^R_\pv(z,z,\omega) \right] =\nn\\
&& 2\gamma (\zv \times \vv_\pv) \Im m \sum_k  \frac{\langle \varphi_k \vert V'(z)\vert \varphi_k \rangle}{(\omega - \epsilon_\pv(k)+i\eta)^2}\,,
\eer
where orthonormality of the states $\varphi_k(z)$ has been used.
A crucial observation is that $-V'(z)$ is the operator of the force exerted by the interface on the electron.
Its expectation value in the scattering state $|\varphi_k\rangle$ is therefore the negative of the pressure exerted by the electron of incoming perpendicular momentum $k$ being reflected at the interface with outgoing perpendicular momentum $-k$.  This pressure is simply the current of perpendicular momentum impinging on the surface, $\frac{2k^2}{m}$.  Thus we have
\be
 \langle \varphi_k \vert V'(z)\vert \varphi_k \rangle =  \frac{2k^2}{m}\,,
 \ee
regardless of the detailed form of the potential.  Armed with this result, we evaluate the momentum sum in Eq.~(\ref{Base}). After an integration by parts we find
\be
\Im m\sum_k  \frac{\langle \varphi_k \vert V'(z)\vert \varphi_k \rangle}{(\omega - \epsilon_\pv(k)+i\eta)^2} =  2\pi\sum_k\delta(\omega - \epsilon_\pv(k))\,.
\ee
Plugging this   into   Eq.~(\ref{Base}) and then Eq.~(\ref{Base})  into Eq.~(\ref{SSD}), and reinstating physical units, we arrive at the promised ``universal" result of Eq.~(\ref{MainResult}),
where the three three-dimensional particle current density $\jv$ is given by $\jv = n\vv_d$, and $n$ is the electron density.

{\it Ab initio calculation for Au(111) surface} -- To demonstrate the usefulness 
of our theorem we have performed an {\it ab initio} atomistic calculation of the spin-density 
profile induced by a current at the Au(111) surface 
-- a system that was analyzed earlier in Ref.~\cite{Krasovskii2011}. The calculations were done for a finite-thickness slab of 19 atomic layers: the self-consistent 
(within the local density approximation) band structure was obtained with augmented 
plane waves method using the full-potential scheme of Ref.~\cite{Krasovskii1999}, and the spin-orbit coupling was included with the second variation
technique of Koelling and Harmon \cite{KOH77}.  The spin-resolved band
structure was calculated in the $\bar\Gamma\bar M$ direction, and, to
simplify the integration over the two-dimensional (2D) Brillouin 
zone, the hexagonal surface was assumed to be axially symmetric.  In the
finite-thickness slab formalism the eigenfunctions are two-component
spinors labeled by a 2D Bloch wave vector ${\bf p}$
parallel to the surface and by a band index $n$, which subsumes the
perpendicular-to-surface component of the Bloch wave vector.  Each Bloch function
contributes a spin density $\frac{\hbar}{2}{\mathbf s}_n({\bf r},{\bf p})$, 
and we are interested in its in-plane $y$ component $s_y({\bf r})$,
perpendicular to the current ${\mathbf j}=\hat{\bf x}j_x$. To
calculate $s_y({\bf r})$ we populate the electronic states with
electrons according to a Fermi distribution shifted by an amount
$\delta{\bf p}=\frac{4\pi e^2}{\hbar\omega_p^2}{\bf j}$, where $\omega_p$
is the plasma frequency entering the Drude conductivity $\sigma =
\tau\omega_{\rm p}^2/4\pi$ 
\footnote{The plasma frequency was taken to be $\omega_p=9$~eV,
as calculated in Ref.~\cite{Glantschnig12010}}.  
After the angular integration (assuming  axial symmetry) we get
\begin{equation}\label{isd}
s_y({\mathbf r}) = j_x\frac{e^2}{2\omega_{\rm p}^2}\sum\limits_{n} s_n\left({\bf r},p^{\rm F}_{ n}\right)p^{\rm F}_{ n},
\end{equation}
where $p^{\rm F}_n$ is the Fermi wave vector in the 2D band of index $n$. 

\begin{figure}
\begin{center}
\includegraphics[width=0.90\columnwidth]{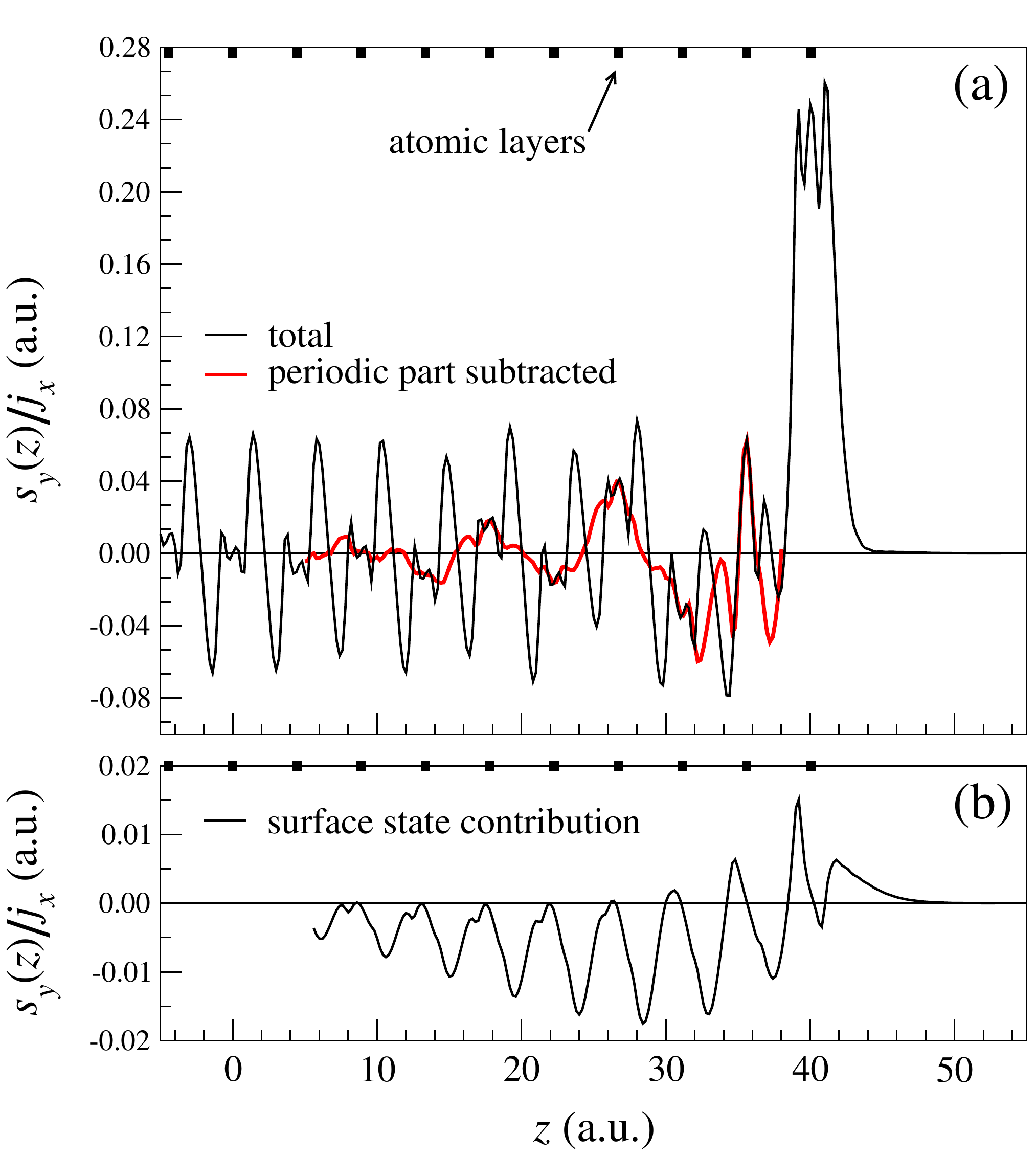}
\caption{\label{induced_sd} (Color online) Spatially resolved 
induced spin density $s_y(z)$ per unit particle current density $j_x$ 
at Au(111) calculated for a symmetric 19-layer slab: (a) total 
spin density and (b) surface state contribution. The center of 
the slab is at $z=0$. The three central layers are seen to be almost 
identical, which proves that the convergence with respect to the layer 
thickness has been achieved. The red curve shows $s_y(z)$ 
with the periodic asymptotic part subtracted.}
\end{center}
\end{figure}

Figure~\ref{induced_sd}(a) shows the depth profile $s_y(z)$, which is the average of  $s_{y}({\mathbf r})$  over the in-plane unit cell.  The function $s_{y}(z)$ has a strong peak  
on the last atomic layer, and deep inside the slab it converges 
to a lattice-periodic function whose integral over the unit cell is zero.  
This spatially-dependent spin polarization, first observed in Refs.~\cite{Kimura2010,Krasovskii2011}, 
arises from the asymmetric occupation of the bulk Bloch periodic 
states in the presence of a current $j_x$. 
The net spin polarization $S_y$ {(integrated over the unit cell)}  must be zero 
due to inversion symmetry ($j_x$ is odd under inversion, $S_y$ is even), 
but a spin-dipole density can and does appear in each unit cell, reflecting the intrinsic 
spin Hall effect of the material.  
This bulk effect is completely absent in the jellium model.
 The red curve in Fig.~\ref{induced_sd}(a) is
obtained from the total $s(z)$ by subtracting 
the lattice-periodic asymptotic function.  {This is the proper surface-induced spin polarization to be compared with the jellium-model calculations.} It is thus seen that the effect of the surface extends over  several atomic layers into the interior of the crystal.\footnote{We note that at the slab thickness of 19 layers the calculations 
are well converged with respect to this parameter.}  
The calculations suggest that the main contribution to the
surface-induced spin polarization comes from bulk-continuum states,
which are not spin-split (they are Kramers degenerate). Somewhat
unexpectedly, the contribution from the Rashba-split surface state is
found to be an order of magnitude smaller, see
Fig.~\ref{induced_sd}(b), {and opposite to the bulk spin polarization}.  The integral surface spin density $S_y$ is
related to the current $j_x$ via Eq.~(\ref{MainResult}) where the
parameter $\gamma$ is obtained by averaging Eq.~(\ref{isd}) over the
2D unit cell and integrating over $z$.

{Expressed in Hartree atomic units (a.u.), for Au(111) the present 
calculation yields $\gamma=0.7\,$a.u. (1 a.u. 
of time is $\hbar/1{\rm H}\simeq 2.42 \times 10^{-17}$~s). Equivalently, using the standard value $m=1.1\,m_e$ of the effective mass for Au  in Eq.~(\ref{gammalambda}), we obtain $\lambda_c^2 \simeq 0.8$ \AA$^2$.} This value can now be used in the effective surface spin-orbit Hamiltonian Eq.~(\ref{HSO}) to reproduce, in the jellium model, the spin
polarization obtained from the {\it ab initio} calculation.  For a charge
current density of $10^{10}$ A/m$^2$ (corresponding to a
particle current density of $10^{29}$ m$^{-2}$s$^{-1})$ we find an induced
spin density of the order of $10^{12}$ m$^{-2}$ (in units of $\hbar$).  It should be possible to observe this surface spin by Kerr rotation microscopy used e.g. to detect spin polarization in 2D electron gases in semiconductors \cite{Kato04-2,Sih05}

Besides parameterizing the spin dependent surface scattering, the value of $\gamma$ will also be useful in modeling bulk effects of great interest in spintronics, such as spin diffusion and the spin Hall effect, particularly when a high degree of accuracy is not required.  In fact, from our {ab initio} $\gamma$ (or equivalently $\lambda_c$) the parameter $\frac{\lambda_c^2 k_F^2}{4}$ which controls the extrinsic spin Hall effect~\cite{Takahashi2008} is found to be $\simeq 0.2$ , which is close the value fitted to experimental transport coefficients for Au (see Table~I in Ref.~\onlinecite{Takahashi2008}).


{\it Acknowledgments} -- This work was supported  by the 
Spanish Ministry of Economy and Competitiveness MINECO
(Projects No. FIS2013-48286-C2-1-P and FIS2013-46159-C3-1-P). I.V.T. acknowledges funding by the Grupos Consolidados UPV/EHU del Gobierno Vasco (Gant No. IT578-13). GV acknowledges support from  NSF grant DMR-1104788 and by the Donostia International Physics Center.

\bibliography{ISO20141001.bib}

\end{document}